\documentclass[sigconf, manuscript]{acmart}

\renewcommand\footnotetextcopyrightpermission[1]{} 
\usepackage{subcaption}
\AtBeginDocument{%
  \providecommand\BibTeX{{%
    \normalfont B\kern-0.5em{\scshape i\kern-0.25em b}\kern-0.8em\TeX}}}

\acmConference[]{}{}{}

\begin{document}

\title{Deep Learning for Enhanced Scratch Input}

\author{Aman Bhargava}
\email{aman.bhargava@mail.utoronto.ca}
\affiliation{%
  \institution{University of Toronto}
  \streetaddress{27 King's College Circle}
  \city{Toronto}
  \state{Ontario}
  \country{Canada}
  \postcode{M5S 1A1}
}

\author{Alice X. Zhou}
\email{alicexyz.zhou@mail.utoronto.ca}
\affiliation{%
  \institution{University of Toronto}
  \streetaddress{27 King's College Circle}
  \city{Toronto}
  \state{Ontario}
  \country{Canada}
  \postcode{M5S 1A1}
}

\author{Adam Carnaffan}
\email{adam.carnaffan@mail.utoronto.ca}
\affiliation{%
  \institution{University of Toronto}
  \streetaddress{27 King's College Circle}
  \city{Toronto}
  \state{Ontario}
  \country{Canada}
  \postcode{M5S 1A1}
}

\author{Steve Mann}
\email{mann@eyetap.org}
\affiliation{%
  \institution{University of Toronto \& MannLab Canada}
  \streetaddress{27 King's College Circle}
  \streetaddress{330 Dundas St. W}
  \city{Toronto}
  \state{Ontario}
  \country{Canada}
}


\begin{abstract}
The vibrations generated from scratching and tapping on surfaces can be highly expressive and recognizable, and have therefore been proposed as a method of natural user interface (NUI). Previous systems require custom sensor hardware such as contact microphones and have struggled with gesture classification accuracy.

We propose a deep learning approach to scratch input. Using smartphones and tablets laid on tabletops or other similar surfaces, our system achieved a gesture classification accuracy of 95.8\%, substantially reducing gesture misclassification from previous works. Further, our system achieved this performance when tested on a wide variety of surfaces, mobile devices, and in high noise environments.

The results indicate high potential for the application of deep learning techniques to natural user interface (NUI) systems that can readily convert large unpowered surfaces into a user interface using just a smartphone with no special-purpose sensors or hardware.
\end{abstract}

\ccsdesc[300]{Human-centered computing~Gestural input}
\ccsdesc[500]{Human-centered computing~Sound-based input / output}
\ccsdesc[300]{Human-centered computing~Accessibility technologies}
\ccsdesc[300]{Human-centered computing~HCI theory, concepts and models}
\ccsdesc[300]{Human-centered computing~User interface toolkits}

\keywords{scratch input, natural user interface, reality user interface, machine learning, deep learning, gestural interface, audio classification}


\begin{teaserfigure}
  \includegraphics[width=\linewidth]{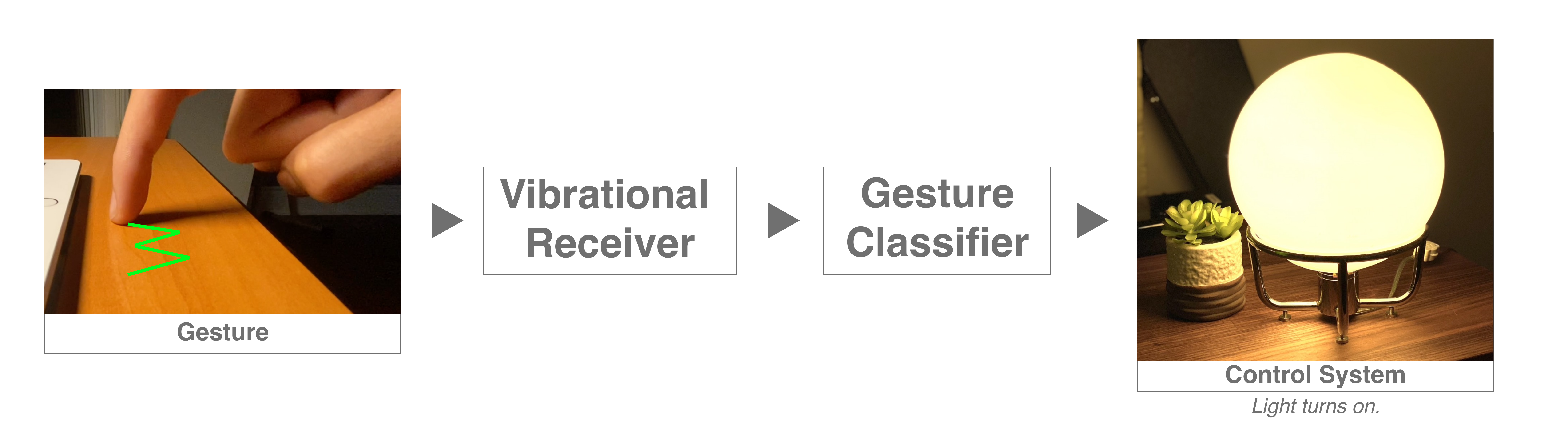}
  \caption{Generalized flow diagram for a scratch input system. Note that multiple vibrational receivers and control systems may be used, assuming that the gesture classifier is correctly configured.}
  \Description{Generalized scratch input system flow diagram.}
  \label{fig:info_flow}
\end{teaserfigure}

\maketitle


\section{Introduction}

There is substantial information, both frequency and time-based, contained in scratches, taps, and swipes made on a textured surface. One can easily observe that the type of gesture, speed, intensity, and shape significantly alter the sound produced. While other factors such as different surface materials and background noise alter the sounds, most gestures remain recognizable.

Scratch input offers the possibility of rapidly deployable, low cost, large scale touch input surfaces. With a sufficiently high-performance system, a wide range of environments could be configured on an ad hoc basis to become an immersive, expressive user interface. Nearly any ``scratchable'' surface could potentially become an interface, producing a form of natural user interface augmented reality (Figure \ref{fig:info_flow}). Potential use cases include home automation, accessibility technology, and user interface in hazardous environments (discussed in Section 1.2).

\subsection{Prior Work on Scratch Input}

This expressiveness of scratching, tapping, and swiping surfaces is likely what inspired the first documented instance of scratch input \cite{mann_physiphone}. Mann \textit{et al} initially proposed scratch input as a source signal and modulation trigger for a novel musical instrument known as a ``physiphone'' or a ``hyperacoustic instrument'' \cite{mann_hyperacoustic}. Shortly thereafter, scratch input was proposed in a human-computer interaction system by Chris Harrison in 2008 \cite{harrison_scratch}. 

In this work, a custom sensing apparatus made of a microphone and a modified stethoscope was used in low-noise environments to classify a set of 6 gestures. Based on energy peak counting and amplitude variation, a hand-written shallow decision tree was used for classification. The gestures investigated all differed clearly in terms of number of energy peaks (e.g., single tap, double tap) or amplitude variation (i.e., scratch vs. tap). To evaluate the effectiveness of the system, users were given five minutes to practice the gestures as they received real-time feedback on the quality of their gestures from the system. Experimenters also provided advice during this time on how participants should adjust their gestures for maximum accuracy. Finally, the 15 participants were requested to perform each of the 6 gestures 5 times each in random order as gesture classification accuracy was recorded.

The study yielded an overall accuracy of 89.5\%, with lower accuracy for complex gestures (triple swipe, quadruple swipe). The investigation offered an overview of potential applications of scratch input including device enclosures and clothing fabric, though the noise in such environments was noted as a challenge for the practical application of scratch input technology. 

Further work has been done by Harrison and others on similar topics since, including investigations on the human body as a transmission medium and input surface in 2010 called \textit{Skinput} \cite{skinput}. In this work, Harrison \textit{et al} successfully localize finger taps on the arm and hand into 20 areas. 186 features were computed based on signal amplitude, standard deviation, Fourier transform, and total energy from each of the 5 receivers in the arm band. Using a support vector machine, a mean classification accuracy of 87.6\% was achieved. In 2011, Harrison introduced \textit{TapSense}, a system to enable the identification of the object being used to tap a surface \cite{tapsense}. A modified stethoscope similar to the apparatus of \cite{harrison_scratch} was used for sensing. Using a support vector machine with 559 features extracted from a Fourier transform computed on a sliding 43ms window, the system achieved roughly 95\% accuracy for differentiating fingertip, fingernail, knuckle, and finger pad taps. Similar methods were used by Ono \textit{et al} in 2013 to recognize different types of touches and grasps on objects using a paired vibration speaker and piezo-electric microphone \cite{touch_and_activate}. A support vector machine with 400 features extracted from a windowed Fourier transform was used to recognize a variety of object grasp positions. 

The general progression in user input and gesture recognition based on vibrations has been toward making use of progressively higher dimensionality features and more complex classifier algorithms. Naturally, questions arise on the applicability of deep learning techniques in such systems to enhance classification accuracy.

\subsection{Potential Use Cases}

In addition to presenting a highly accurate scratch input gesture classifier, we highlight the utility of the proposed system for ``internet of things'' (IoT) implementations in home automation, for human-computer interaction accessibility, and for use in harsh environments where conventional computer interaction methods are infeasible (e.g., front-line workers in biohazardous and other extreme environments).

\subsubsection{Home Automation}

The proposed system could readily be used in a variety of home automation use cases. High-accuracy user input on a relatively small set of gestures is readily applied to control tasks in home automation \cite{home_auto_gestures}. Advantages of the proposed scratch input system include the rapid deployment of extremely large, low-cost interface surfaces and environments. With several low-cost microphones or networked mobile devices, any space could be converted into an interactive, controllable, and responsive environment. 

Since any scratching or tapping gesture can be used for the system that consistently differs from the rest acoustically, gestures could also be made intuitive and language-independent. For example, the \textbf{vertical scratch} and \textbf{circular scratch} gestures, resembling a binary `1' (on) and `0' (off) respectively, may be used to turn on and off a lamp. Similarly, the \textbf{W-scratch}, resembling iconographic depictions of water \cite{water_icon}, may be used for faucets, flushing, or other related functions.

When compared with existing commonly sold voice assistants for home automation, the proposed system also stands out in terms of privacy. All computation for gesture classification and noise discrimination could easily be implemented onboard with a low-cost microprocessor or system on a chip. Meanwhile, voice assistants today are nearly universally ``cloud based'', entailing a host of security concerns \cite{voice_assistant_security}. To execute commands, voice recordings are recorded and sent to a remote system (``cloud'') for processing before execution. In production, the proposed deep learning-based scratch user interface system could easily separate the gesture classification system from the networked components, drastically increasing user privacy and security.

\subsubsection{Accessibility}

As addressed in \cite{harrison_scratch}, the majority of computer systems in common use today are relatively small for ease of transportation (e.g., mobile phones, laptops). Even in the home, where it may be possible to deploy a larger scale user interface system, it is often inconvenient and prohibitively expensive. This problem is especially important for those people affected by disabilities relating to pain, flexibility, and dexterity. According to a 2012 Canadian Survey on Disability, over 25\% of those reported to have some type of disability were identified as having  dexterity disability that limited their daily activity \cite{disability_survey}. 

A natural user interface (NUI) system that could easily expand to the size of a table could help to ameliorate these limitations on daily activity as they relate to the use of ubiquitous computing systems. As will be discussed later, the proposed system could also be configured to continuously learn and improve on a user's data, enabling adaptation to each individual's accessibility needs. Since the proposed system was validated and tested with a wide variety of consumer smartphone and tablet microphones, it would be rapidly deployable and very likely low-cost for the end user given the widespread use of smartphones.

\subsubsection{Hazardous Environments}

Given the proposed system's ease of use for system control tasks like home automation in with little to no reliance on dexterity, it is logical to propose its use in extreme environments. In areas of extreme cold, for instance, thick protective gear must be worn that severely inhibits manual dexterity. While conventional computer system input devices may be difficult to use, scratches and taps remain easy to replicate. A set of gestures optimized for use with protective equipment could be easily designed and implemented with the following system, aiding in the maintenance of safety and efficiency as safety equipment would need not be removed to use the system.

\subsection{Design Goals, Scope, and Principles}

\begin{figure}[h]
  \centering
  \includegraphics[width=\linewidth]{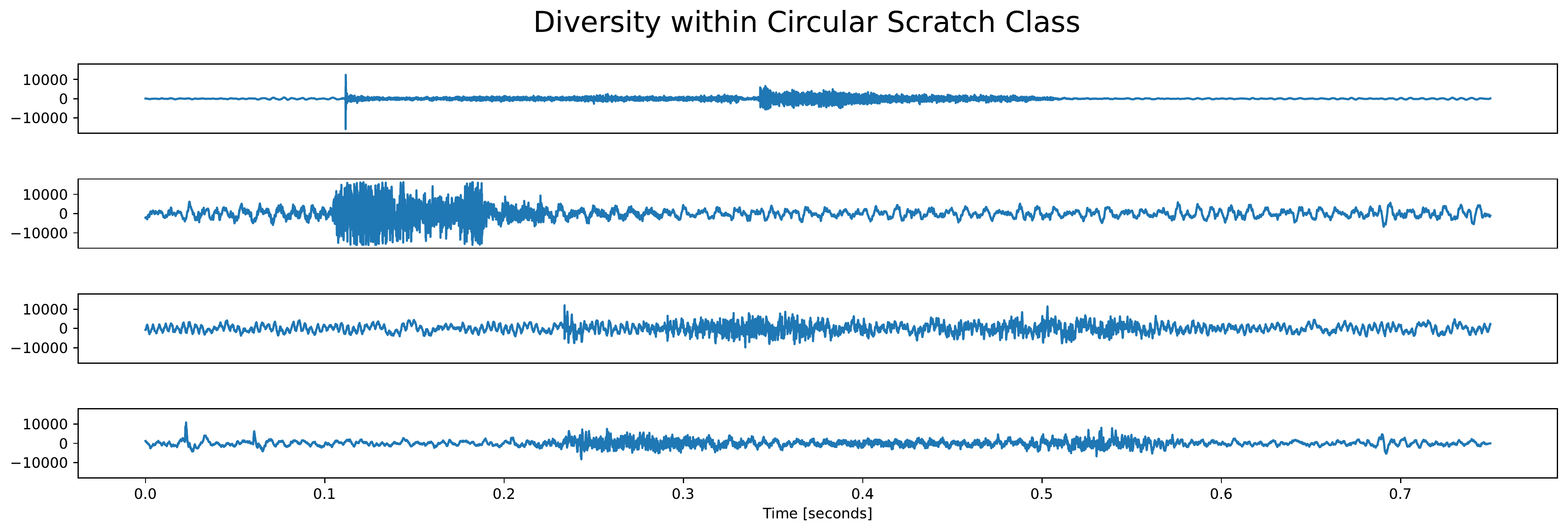} 
  \caption{Examples waveforms from a variety of users, devices, and environments from the \textbf{circular scratch} gesture class. Note how background noise, amplitude, speed, and shape differ substantially in real-world use cases.}
  \Description{Examples waveforms for each input class.}
  \label{fig:speed_amp}
\end{figure}

Deep learning and machine learning are a natural step forward for scratch input technology. With enough data on a well-defined problems, these learning algorithms are often able to perform extremely well on hitherto unsolved automated classification and regression problems such as object recognition, speech recognition, and natural language processing. This investigation seeks to understand the extent to which deep learning techniques can be used to enhance scratch input technology. We seek improvement in terms of the following metrics:
\begin{enumerate}
    \item Increased gesture classification accuracy.
    \item Decreased reliance on custom apparatus.
    \item Increased performance with a diversity of users and noisy environments on similar-sounding gestures (Fig.~\ref{fig:speed_amp}). 
\end{enumerate}

To investigate this, a dataset of over 8,000 audio gesture clips were collected using generic mobile devices by a range of users in diverse environments. The proposed deep convolutional neural network achieved 95.8\% testing accuracy despite increased noise, gesture similarity, and surface diversity. Several ``baseline'' machine learning models utilizing waveform representation techniques similar to \cite{google_audio} were also trained to validate performance gains made by the deep learning approach.

\section{Proposed System}

\begin{figure}[h]
  \centering
  \includegraphics[width=\linewidth]{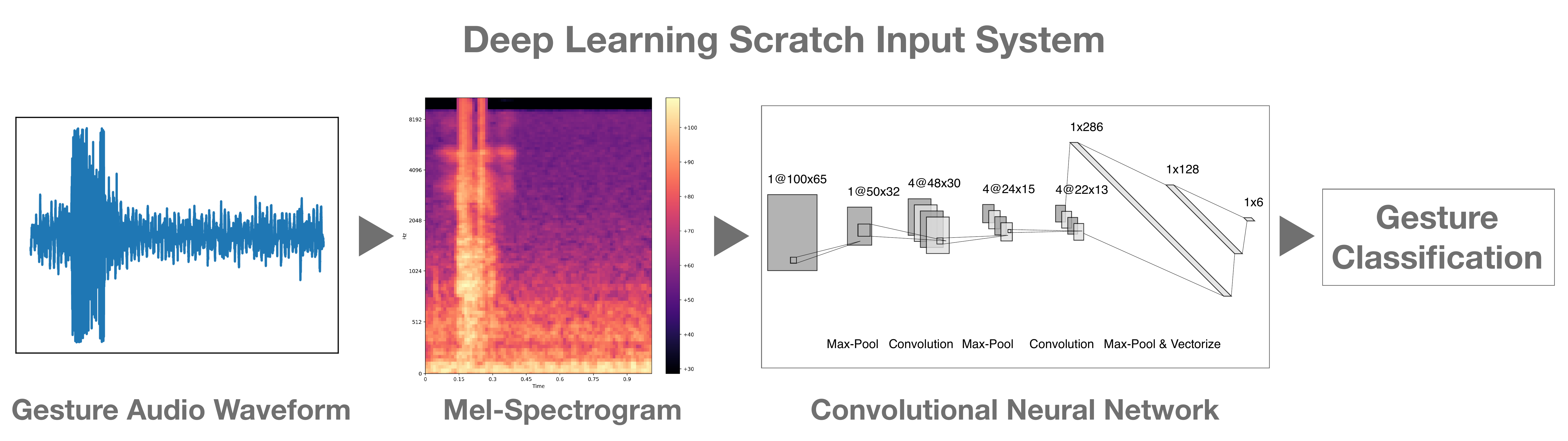}
  \caption{Flowchart of the proposed system utilizing Mel-spectrograms for feature extraction and a convolutional neural network for classification.}
  \Description{Generalized scratch input system flow diagram.}
  \label{fig:prop_sys}
\end{figure}

We propose a high-performance audio gesture processing system (Figure \ref{fig:prop_sys}) that uses a convolutional neural network (CNN) for classification in conjunction with a Mel-spectrogram. In order to test the applicability of the system, a dataset of 8,640 audio samples was collected, processed, and used to train and test the proposed system and two baseline models. Full code for data processing, feature extraction, model architecture, training, and validation can be found in the project \href{https://github.com/xxxx/xxxx}{GitHub repository (link)}.

\subsection{Data Collection}

The dataset for this experiment was evenly split between 6 gesture classes:
\begin{enumerate}
    \item \textbf{Fingertip Tap}
    \item \textbf{Fingernail Tap}
    \item \textbf{Vertical Scratch}
    \item \textbf{Circle Scratch}
    \item \textbf{W-Scratch}
    \item \textbf{Silence/null class}
\end{enumerate}


This gesture set was chosen so that results obtained would be comparable to those in \cite{harrison_scratch} where 6 gesture classes were also tested. Video instructions on gathering the data were sent to 24 participants, each with different surfaces, environments, and mobile devices for recording. The video provided instructions and examples for how to execute each gesture along with a metronome recording to help users achieve consistent timing between gestures. Each participant recorded 70 audio gesture segments of each input classes. During the \textbf{silence/null} class recording, participants were encouraged to do other work and activities, resulting in realistic background noise (e.g., typing, page rustling, talking). To clean mis-timed and improper audio gesture segments, the first 10 gestures from each recording were removed from the dataset, resulting in a final dataset of 8,640 audio gesture segments. 

Recordings contained both steady-state and transient background noise. Gesture recordings varied substantially in amplitude and speed. Different mobile device models also varied in terms of frequency filtering, microphone frequency response, and recording bitrate. 12 of the 24 participants used iPhones or iPads from 2017 or later while the rest used devices produced by Huawei, Xiaomi, Samsung, Pocophone, Google, Motorola, LG, or One Plus in a similar time period. The full dataset with device specifications is available from the authors upon request.

\subsection{Data Processing $\&$ Feature Extraction}

\begin{figure}[h]
  \centering
  \includegraphics[width=\linewidth]{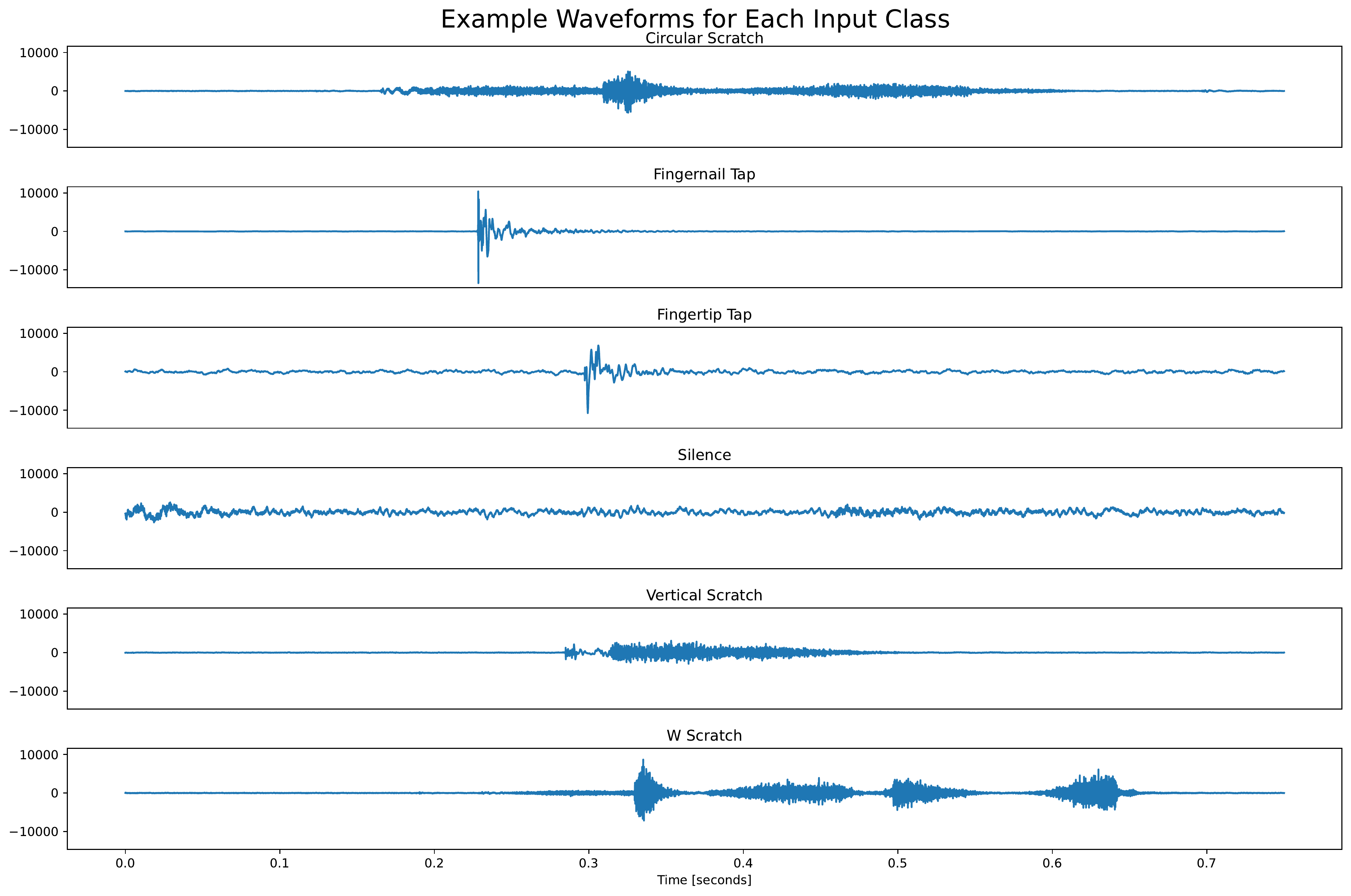}
  \caption{Examples waveforms for each input class.}
  \Description{Examples waveforms for each input class.}
  \label{fig:wave_classes}
\end{figure}

The raw audio files were manually cropped and aligned such that the gestures all occurred at the same times. Audio was then re-sampled at 44.1 kHz. The clips were segmented into individual gesture recordings, resulting in a dataset of 8,640 individual gesture audio files, each between 0.75 and 1.5 seconds long. For training and testing purposes, each gesture was symmetrically cropped to the minimum length of 0.75 seconds (see Fig.~\ref{fig:wave_classes}).

To represent the waveforms to the machine learning algorithm, additional feature extraction was required. If the audio recordings were directly inputted, each audio vector would have dimensionality on the order of $10^4$ due to the 44.1 kHz re-sampling rate. Past studies have confirmed that deep learning on such feature vectors is incredibly computationally expensive and tends to lead to poor results (overfitting) \cite{audio_overfit} \cite{palanisamy2020rethinking}. 

The feature extraction method chosen for this study was to convert the audio gesture recordings into Mel-frequency spectrograms. Mel-frequency spectrograms are a type of spectrogram with a logarithmic frequency and amplitude scale \cite{zwicker1961subdivision}. They are produced by generating Fourier transforms of a signal on short time windows and calculating the energy contained in each logarithmically-spaced frequency interval (known as a ``Mel''). 

This technique has yielded excellent results in state-of-the-art deep learning audio classification studies \cite{google_audio} \cite{thorntonaudio} \cite{9349416} due to its favourable representation of the time-frequency attributes of a signal. Because of the logarithmic frequency scale, an increased or decreased fundamental frequency simply results in vertical translation on the plot without the dilation or contraction associated with typical spectrograms. Thus, differently pitched scratches and taps tend to differ only in location on the spectrogram, improving the CNN's ability to generalize a rule for recognizing each gesture class. Mel-spectrograms of size 100 by 65 pixels were generated, corresponding to a 0.75 second long rolling ``processing window''.


\subsection{System Design and Training}


The CNN consists of two convolutional layers each composed of a number of 3x3 kernels followed by two fully connected layers with 100 hidden units. 2x2 max pooling was used after each of the convolutional layers, and ReLU activation functions were used throughout. For training, cross entropy loss was minimized via stochastic gradient descent (SGD). Batch normalization was employed \cite{ioffe2015batch} to aid in training. A softmax function was applied to the output of the CNN to obtain class probabilities. A graphical depiction of the CNN can be seen in Fig. \ref{fig:prop_sys}.

Other hyperparameters including SGD batch size, learning rate, and the number of 3x3 kernels per convolutional layer were selected via grid search (see Table \ref{tab:hyper-parameters}). To train the classifiers, the dataset was split into a training dataset (60\%), a validation dataset (15\%), and a testing dataset (25\%). The training set was used in each grid search iteration to train the model  for 50 epochs while the validation dataset was used to select the best model from the grid search procedure. The final accuracies are reported in terms of performance on the separate testing set (per \cite{goodfellow2016deep}, \cite{yaser2012learning}).

\section{Results}

\subsection{Baseline Classifiers}

To validate the utility of computationally expensive deep learning approache over simpler machine learning approaches, two baseline classifiers were trained on the dataset. 

\begin{figure*}[t!]
    \centering
    \begin{subfigure}[t]{0.5\textwidth}
        \centering
        \includegraphics[height=3.5in]{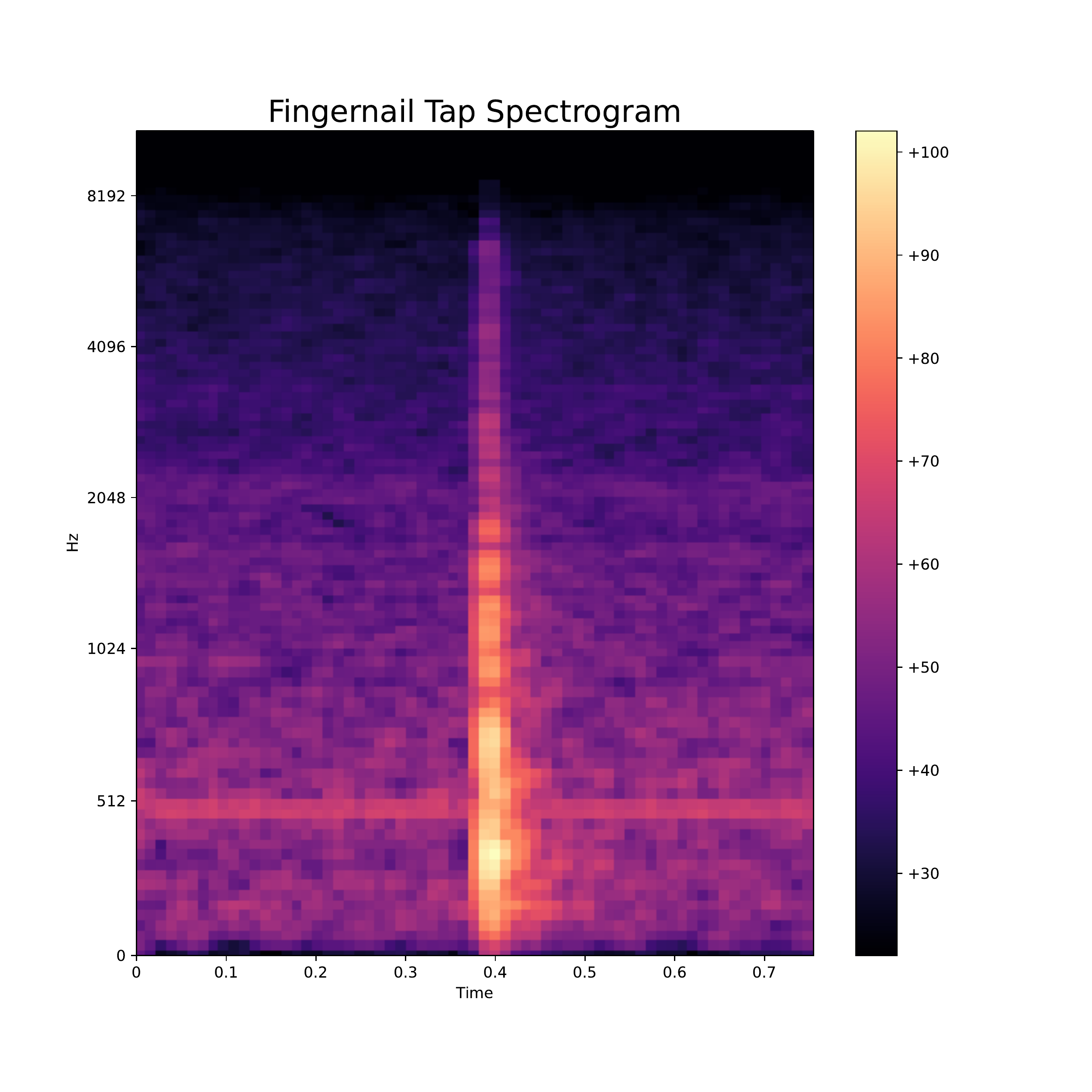}
    \end{subfigure}%
    ~ 
    \begin{subfigure}[t]{0.5\textwidth}
        \centering
        \includegraphics[height=3.5in]{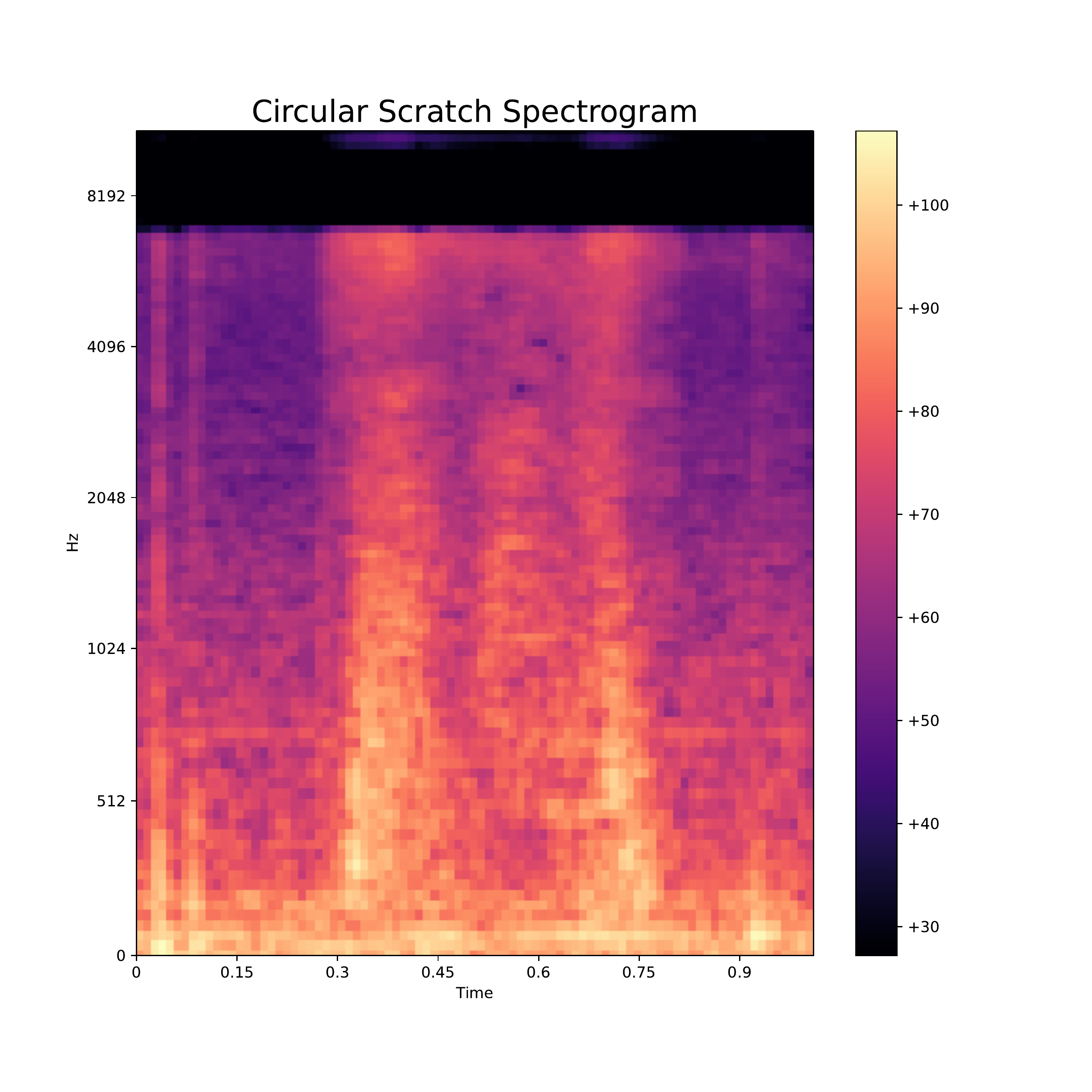}
    \end{subfigure}
    \caption{Example spectrograms from the \textbf{fingernail tap} and the \textbf{circular scratch} classes.}
    \label{fig:mel_examples}
\end{figure*}

The first baseline classifier was trained using feature vectors representing signal energy as a function of time similar to \cite{harrison_scratch}. Features for this classifier were extracted by summing along the columns of the Mel-frequency spectrograms (Fig~\ref{fig:mel_examples}), resulting in 65-dimensional feature vectors. A Multi-Layer Perceptron (MLP) model was used with 1 hidden layer with ReLU activation functions. 100 neurons were used in the hidden layer, and with the Adam optimizer, the model achieved a test set accuracy of \textbf{71.2\%}. This accuracy is considerably lower than that in \cite{harrison_scratch} despite the substantially more complicated MLP model, indicating that time-energy features may not be sufficient in higher noise and more more diverse environments for this gesture set.

The second baseline classifier used the same MLP model as the first. The feature vectors in this case represented energy as a function of frequency, similar to \cite{skinput}, \cite{touch_and_activate}. Feature vectors were calculated by summing along the rows of the Mel-frequency spectrograms. The 100-dimensional feature vectors yielded a test accuracy of \textbf{81.2\%}. This result indicates that frequency-energy information may be more helpful for gesture classification in higher noise/variance environments.

\subsection{CNN Model Results}

\begin{figure}[h]
  \centering
  \includegraphics[width=\linewidth]{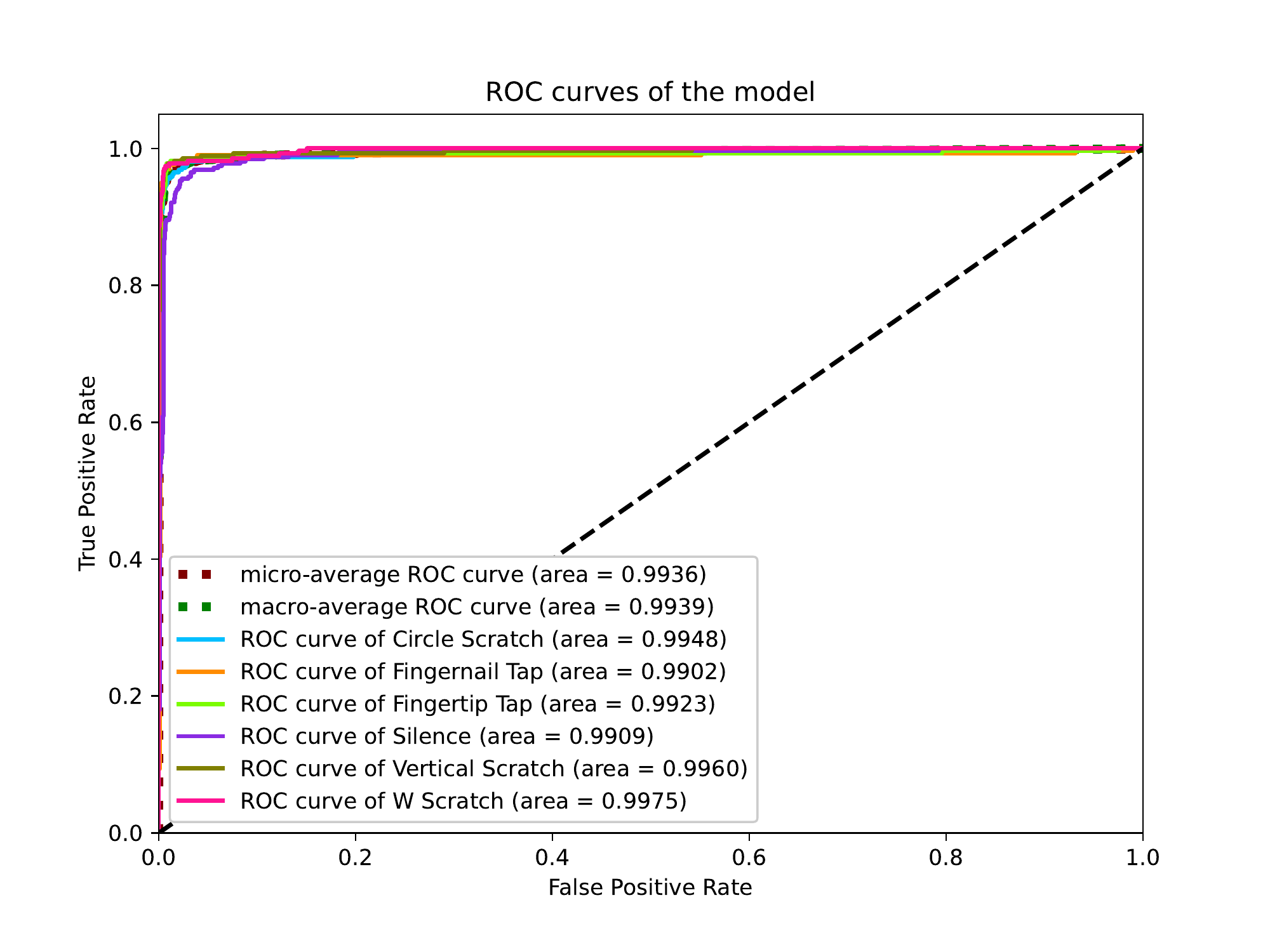}
  \caption{ROC curves for final CNN model on testing dataset.}
  \Description{Plot demonstrating ROC for cropped model}
  \label{fig:roc_nos}
\end{figure}

The overall test set accuracy of the final CNN classifier selected using grid search  \textbf{95.8\%}. Several methods were used to validate the performance of the classifier in addition to overall accuracy on the testing dataset. First, the receiver-operating characteristic (ROC) curves were calculated. ROC curves plot the rate of false positives against true positive rates for different ``certainty cutoffs'' on the model output. In the best case, the ROC curve would pass through the top left corner (Fig. \ref{fig:roc_nos}), indicating that the model could achieve 100\% true positive rate and 0\% false positive rate. The area under the ROC curve is the de facto measurement for classifier success in machine learning, and was calculated for each gesture's classification. An area of 1 represents perfect classification and an area of 0.5 represents random guessing \cite{fawcett2006introduction}.

The ROC curves show extremely high performance on all classes, with the lowest class being \textbf{Fingernail Tap} and the highest being \textbf{W-scratch}, each with area-under-curve value of 0.9902 and 0.9975 respectively (Fig.~\ref{fig:roc_nos}).

A confusion matrix (Table.~\ref{tab:conf_nos}) was also calculated for the model's classification performance on the test set. The horizontal labels at the top represent the algorithm's classification while the vertical labels to the left represent the ground truth class value for each sample in the testing set. The confusion matrix shows that the algorithm most commonly output the \textbf{silence} class when ``uncertain''. The \textbf{fingernail tap} class was most commonly confused for another class.  

\begin{table*}[h]
  \caption{Confusion Matrix for Cropped Test Data}
  \label{tab:conf_nos}
  \begin{tabular}{*{6}{c}l}
    \toprule
     & Circle Scratch & Fingernail Tap & Fingertip Tap & Silence & Vertical Scratch & W-Scratch\\
    \midrule
    Circle Scratch & 302 & 0 & 0 & 4 & 2 & 2 \\
    Fingernail Tap & 2 & 292 & 4 & 5 & 1 & 1 \\
    Fingertip Tap & 3 & 5 & 269 & 3 & 0 & 0 \\
    Silence & 6 & 3 & 4 & 300 & 4 & 3 \\
    Vertical Scratch & 2 & 0 & 0 & 3 & 266 & 1 \\
    W-Scratch & 2 & 0 & 0 & 2 & 1 & 265 \\
    \bottomrule
  \end{tabular}
\end{table*}

During model testing, certain practical qualitative patterns were observed. In particular, it was observed that the \textbf{circle scratch} class was frequently misclassified as \textbf{silence}. 

\begin{figure}[h]
  \centering
  \includegraphics[width=\linewidth]{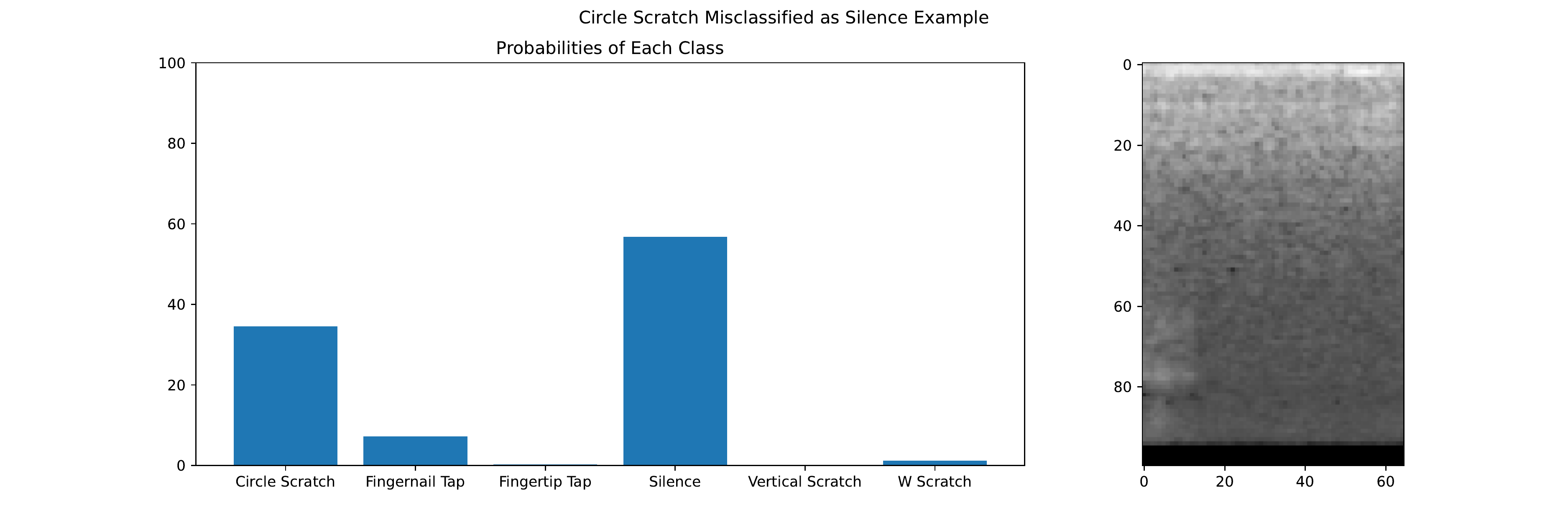}
  \caption{Live Demo Perception of \textbf{Circle Scratch} Input - Confused}
  \Description{A chart of probabilities for different output classes 
  coupled with the image being analyzed}
  \label{fig:ftex}
\end{figure}

Confounding examples of this and other classes showed substantial visual similarity to the improper class (Fig.~\ref{fig:ftex}). In the case of the \textbf{circular scratch}, misclassified gestures often lasted longer than the total length of the spectrogram frame and therefore appeared similar to steady-state background noise instead of a transient scratch. 
 
\section{Discussion}

The high and consistent accuracy from the CNN model shown in Table \ref{tab:conf_nos} and Figure \ref{fig:roc_nos} show that deep learning can be used to enhance scratch input to a great extent. Despite environmental noise resulting from the use of non-contact microphones and variation in environment and surface, the proposed system for scratch input yielded 95.8\% test set accuracy. 

The experiments also raised important questions relating to the practical implementation of such a deep learning scratch input system. Namely, the active time window duration of each segment and the generalizability of the model to a new surface/environment are of interest to characterize the applicability of the findings. Advantages and disadvantages of the deep learning approach as compared to previous works are also of interest. 

\subsection{Comparison to Prior Work}

The use case for the proposed system most closely resembles that of Harrison and Hudson 2008 paper on scratch input \cite{harrison_scratch} with key differences in apparatus, feature extraction, and algorithmic complexity. 

In large part due to the increased algorithmic complexity and lengthy training procedure, the proposed deep learning system was able to achieve a 95.8\% classification accuracy on the 6 gesture types while Harrison and Hudson's system achieved an 89.5\% classification accuracy. The proposed system also achieved this accuracy in spite of additional background noise and variation resulting from the use of miscellaneous phone and tablet microphones rather than the stethoscope apparatus used by Harrison and Hudson. The gestures in the present investigation did not differ in terms of energy profile as cleanly as those of \cite{harrison_scratch}, and the users did not have the opportunity to practice with the system before recording their data.

This increase in classification accuracy comes at a cost, however. The CNN algorithm and the Mel-spectrogram feature extraction in the proposed system is substantially more computationally expensive. It also lacks the interpretability of Harrison and Hudson's shallow decision tree algorithm where one could theoretically trace the genesis of a classification error to either a miscount of the energy peaks or an incongruent amplitude variation measurement. As well, the proposed system requires a ``silence'' class to differentiate null input from other gestures while prior work does not. 

It is clear that a deep learning approach to scratch input has the potential to drastically decrease gesture misclassification for scratch input and related interaction modalities, even in noisy and high-variance environments with diverse apparatus. The system increases classification by a roughly 5\% from past work, corresponding to a decrease in gesture misclassification from 10.5\% (\cite{harrison_scratch}) to 4.2\%, or a 60\% decrease in gesture misclassification. 

\subsection{Window Sizing Tradeoff}

One of the important trade-offs that must be made when used with a spectrogram/convolutional neural network for gesture classification relates to the duration of each gesture recording. In practice, the algorithm has an ``active window'' of audio that it processes at a given time. This corresponds to the width (i.e. duration) of the spectrogram images used as input for the model. For convolutional neural networks, this size must be constant.

For these tests, the duration of each gesture segment was cropped to the exact same length (0.75 seconds). This places a lower limit on the spacing between two subsequent gestures as well as an upper limit on the length (and amount of information available) on each gesture. It was observed that longer gestures such as \textbf{W-scratch} and \textbf{circular scratch} would at times exceed the duration of the frame, resulting in the problem discussed in subsection Section 3.2.

\begin{figure}[h]
\centering
\begin{minipage}{.3\textwidth}
  \centering
  \includegraphics[width=\linewidth]{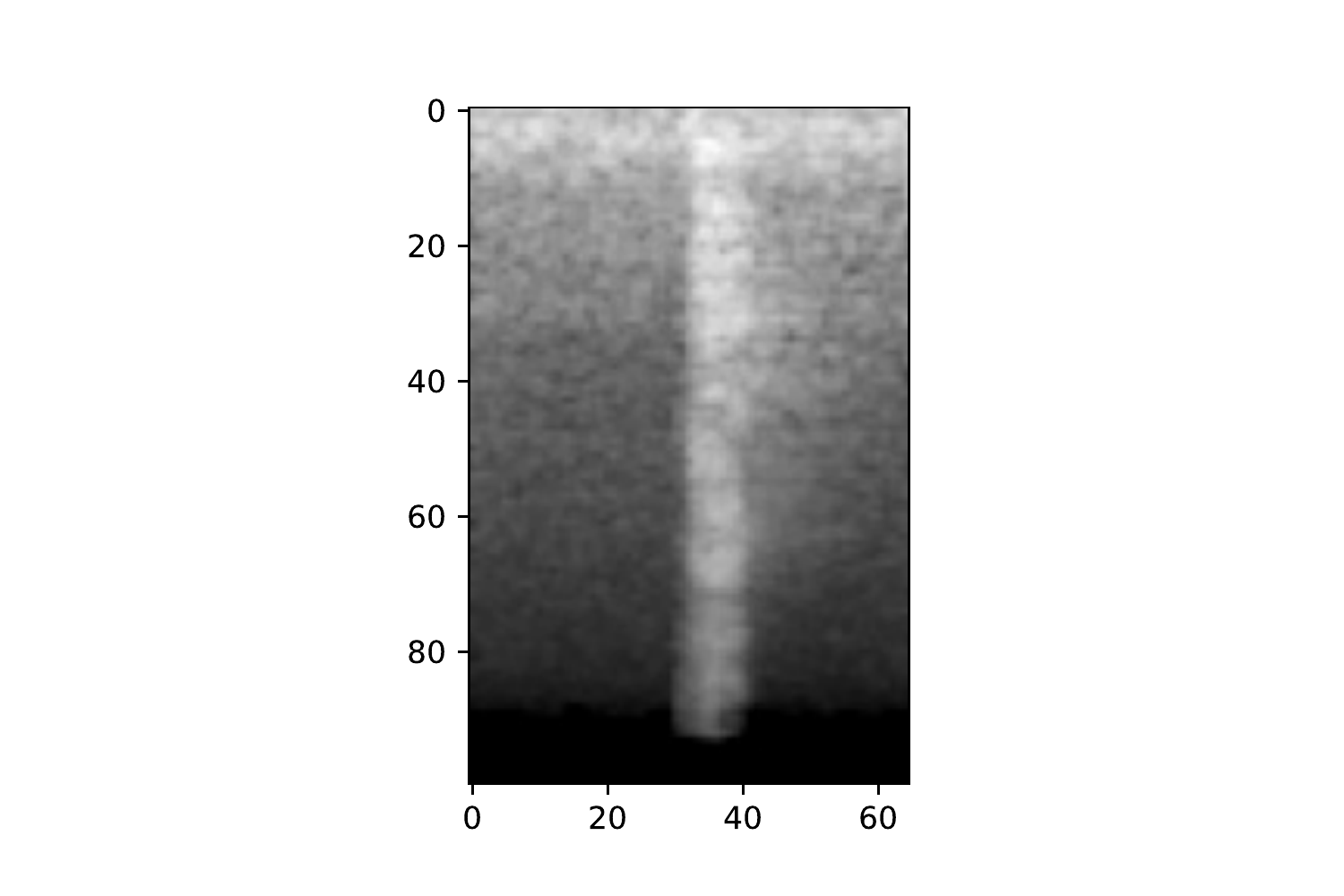}
  \captionof{figure}{Spectrogram image after cropping transformation (\textbf{vertical scratch})}
  \label{fig:crop_vs}
\end{minipage}\qquad
\begin{minipage}{.3\textwidth}
  \centering
  \includegraphics[width=\linewidth]{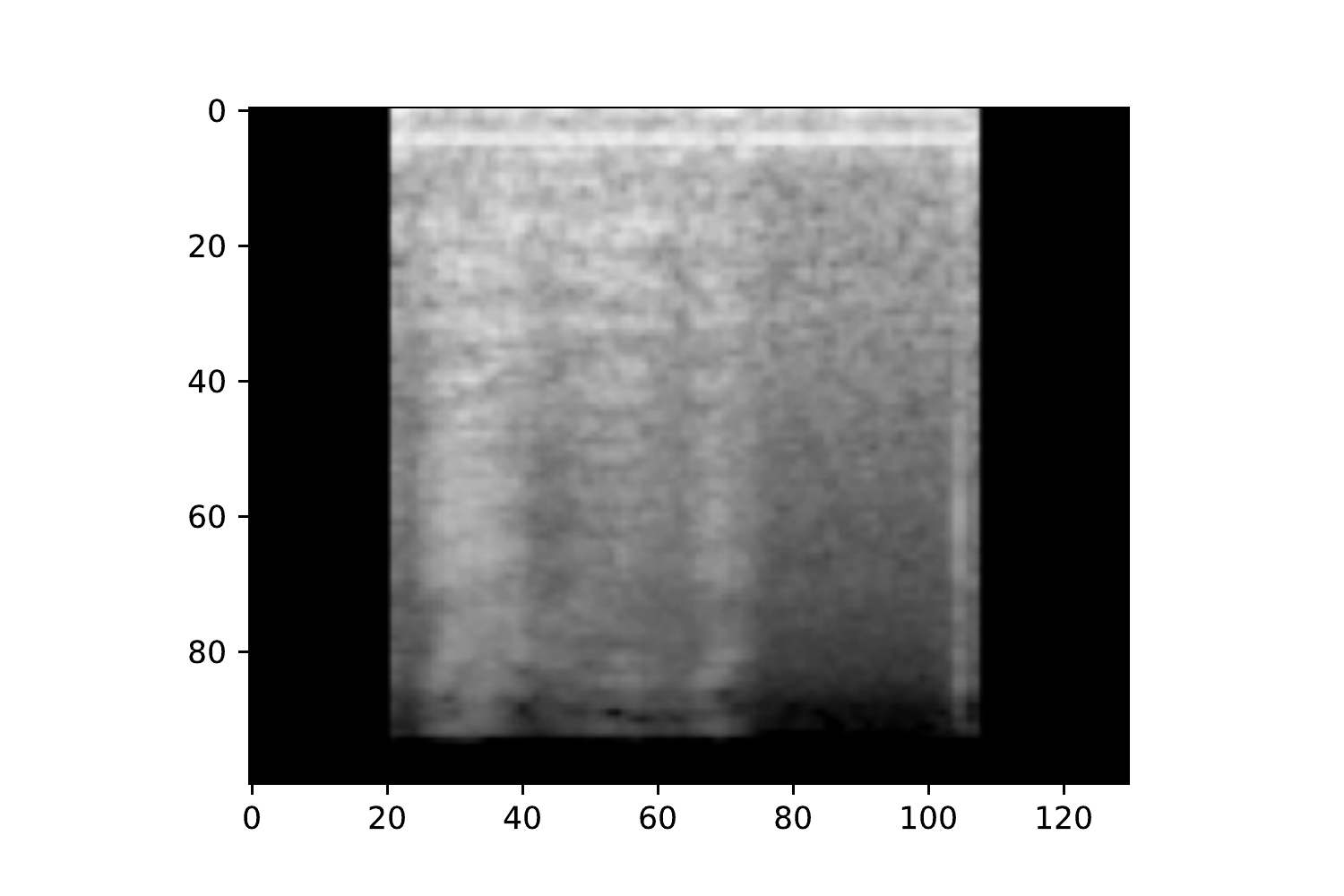}
  \captionof{figure}{Spectrogram image after padding transformation (\textbf{circle scratch})}
  \label{fig:pad_cs}
\end{minipage}\qquad
\begin{minipage}{.3\textwidth}
  \centering
  \includegraphics[width=\linewidth]{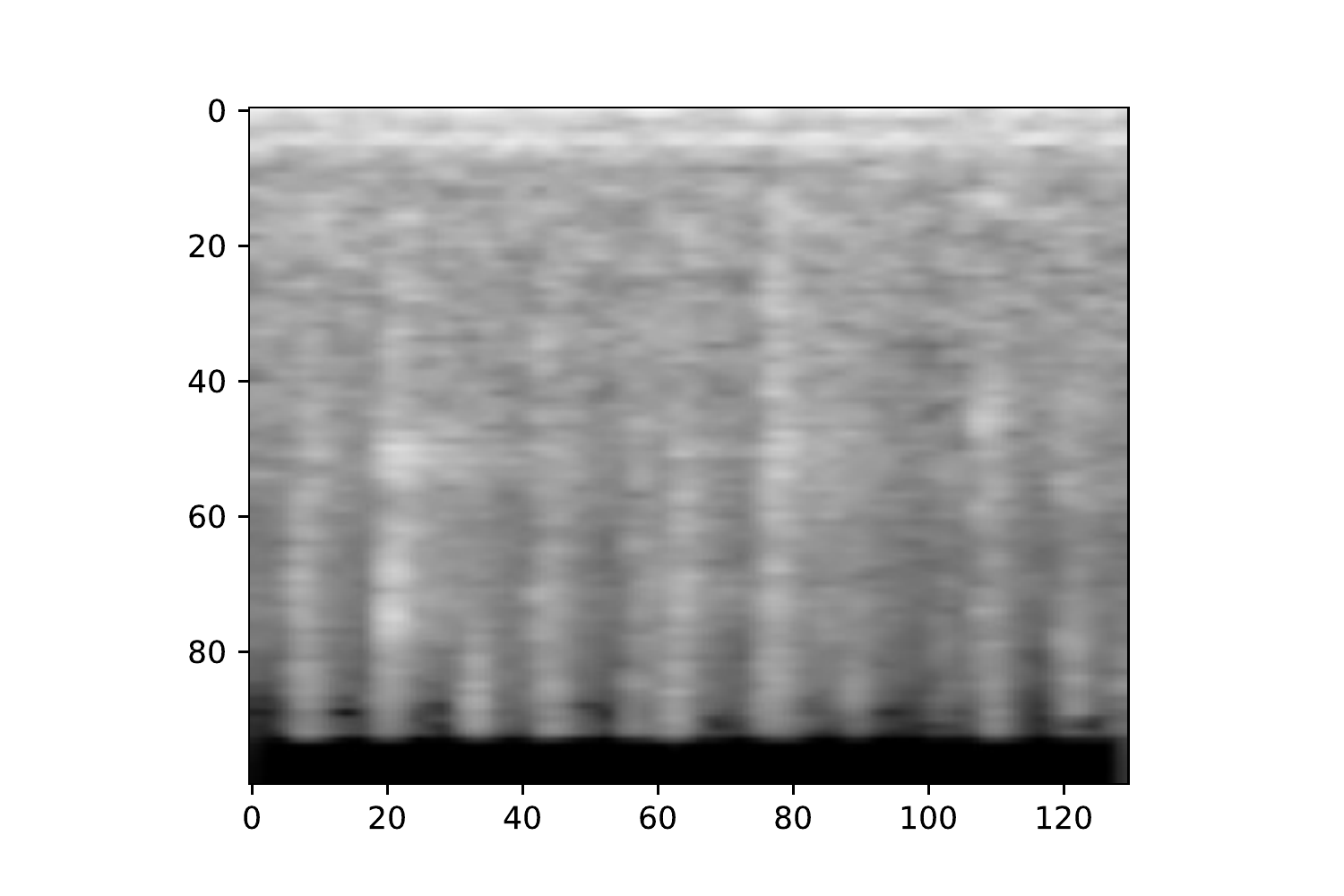}
  \captionof{figure}{Spectrogram image after scaling transformation (\textbf{silence/null})}
  \label{fig:scale_s}
\end{minipage}
\end{figure}

To investigate the amount of information that is lost by cropping longer gestures, multiple alternatives for gesture spectrogram resizing were tested. The original gesture recordings ranged from 0.75 to 1.5 seconds long, there was an opportunity to experiment with scaling up the shorter segments, scaling down the larger ones, or simply padding all of the gestures to a uniform size with black space, as seen in Figures \ref{fig:crop_vs}, \ref{fig:pad_cs}, \ref{fig:scale_s}. While these approaches are not feasible in practice, the results aid in understanding information lost or gained because of the ``active window'' size.

Table.~\ref{tab:best_model} lists the highest accuracies achieved the model using different image transformation methods, with the highest accuracy achieved from padding, lowest accuracy from cropping, and scaling (squeeze/ stretch) in the middle. This result likely stems from information being lost from cropping the images, distorted from scaling, and mostly retained in padding. \\

\begin{table*}[h]
  \caption{Best model accuracies for different image transformations)}
  \label{tab:best_model}
  \begin{tabular}{*{6}{c}l}
    \toprule
     &
    Cropped&
    Padded&
    Scaled Small&
    Scaled Large&
    Scaled Medium\\
    \midrule
    Seed&
    6&
    6&
    6&
    6&
    6\\
    Learning Rate&
    0.3&
    0.1&
    0.1&
    0.1&
    0.1\\
     Batch Size&
    126&
    48&
    48&
    64&
    48\\
    Kernels&
    26&
    4&
    12&
    12&
    8\\
    Test Accuracy&
    95.80\% &
    97.15\% &
    96.41\% &
    96.36\% &
96.07\% \\
    \bottomrule
  \end{tabular}
\end{table*}

In the context of user interface and human-computer interaction, small differences in classification accuracy can substantially change the user's experience. Therefore, meaningful engineering design and research is warranted to determine the optimal window size and gesture set for use in a production system. Investigation on the use of non-windowed machine learning approaches such as recurrent neural networks may also be of value to address these limitations.

\subsection{Calibration Requirements}

Per Section 2.2, the full 8,640-element dataset was randomly split into testing, training, and validation subsets. Thus, each model was trained and tested on every participants data. The same background noise and environments were represented in the training, validation, and testing subsets. This likely made classification easier since the model had already ``seen'' each environment and noise distribution. A production system would not necessarily have this benefit, unless some calibration were used. 

To understand the model's performance in new environments without calibration, the training and testing were re-executed, now split on a subject-by-subject basis. The hyperparameters selected via grid search were re-used for the experiment. The model therefore was not exposed to the particular environment or noise distribution of the subjects in the testing set, emulating a production environment. This resulted in a 76\% overall accuracy and the following results (Fig.~\ref{fig:roc_s}, Table.~\ref{tab:conf_s}):

\begin{figure}[h]
  \centering
  \includegraphics[width=\linewidth]{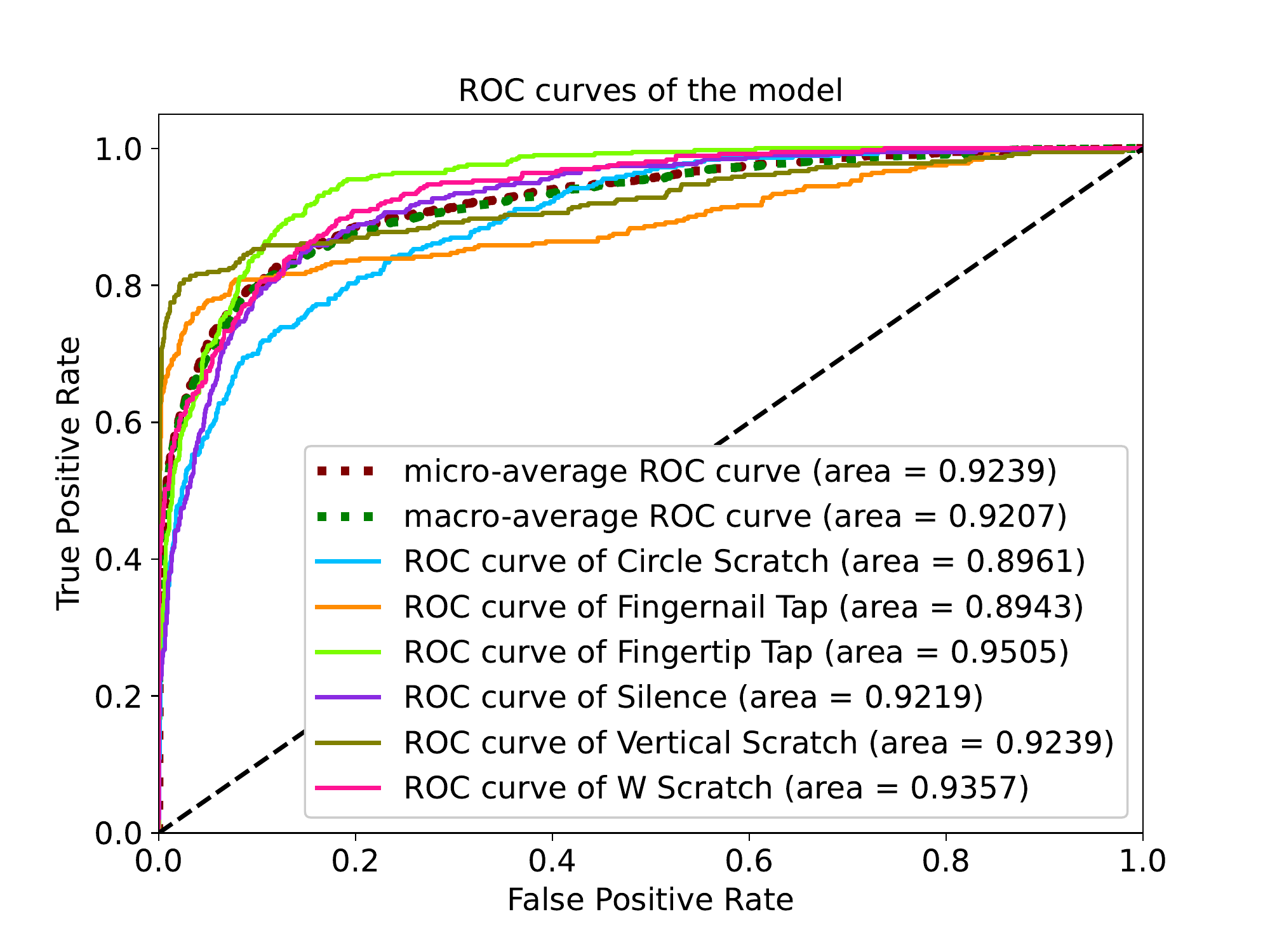}
  \caption{Test set ROC curves for participant-based train/test split model.}
  \Description{Plot demonstrating ROC for cropped model}
  \label{fig:roc_s}
\end{figure}

\begin{table*}[h]
  \caption{Confusion Matrix for Cropped Test Data (From an Unseen Surface)}
  \label{tab:conf_s}
  \begin{tabular}{*{6}{c}l}
    \toprule
     & Circle Scratch & Fingernail Tap & Fingertip Tap & Silence & Vertical Scratch & W-Scratch\\
    \midrule
    Circle Scratch & 261 & 15 & 10 & 24 & 19 & 35 \\
    Fingernail Tap & 2 & 291 & 31 & 1 & 6 & 5 \\
    Fingertip Tap & 4 & 10 & 325 & 32 & 1 & 0 \\
    Silence & 61 & 8 & 16 & 299 & 18 & 38 \\
    Vertical Scratch & 7 & 2 & 1 & 6 & 291 & 10 \\
    W-Scratch & 25 & 34 & 37 & 34 & 25 & 272 \\
    \bottomrule
  \end{tabular}
\end{table*}

Since performance is clearly inferior to the initial results, it would be necessary to include this calibration step in a production system to expose the model to the noise distribution and surface characteristics of a new environment. To emulate the training of the initial 95.8\% accuracy model, only a few minutes would need to be spent to recording the 6 gestures to augment the training dataset. This is a challenge to consider for this deep learning approach when comparing its overall applicability with previous works.

\subsection{Future Work}

One of the primary areas for further exploration relates to the user experience of a deep learning-based scratch input system. Optimization of ``active window'' size and gesture duration would have an outsized effect on user experience. Similarly, closer examination of problematic noises (e.g., pen scratches or typing) and associated mitigation methods are of interest. Power efficient implementation of the processing pipeline would also be important for production implementations of deep learning scratch input.

Another topic of interest is the incorporation of top-down approaches similar to those presented in previous work (e.g., the shallow decision tree used in \cite{harrison_scratch}) with the proposed deep learning methods. As mentioned in Section 4.1, the deep learning approach is not without its disadvantages despite its increased classification accuracy. Interpretability, reliability, and energy concerns may be addressed via a hybrid system that incorporates alternative approaches to scratch input.

The calibration process required to use the deep learning scratch input system with high accuracy is worthy of investigation. In particular, the use of transfer learning and data augmentation could aid in further reducing the time taken to calibrate the system. A natural next step for the deep learning system could also involve continuous learning and improvement as the system is used. 
    
As proposed, the system is a discrete input system with a relatively small set of classes. Incorporating one or more spatial dimension into the input signal could enhance the usability. This could potentially be accomplished via multiple microphones or by using a computer vision system to locate the user's gesture. As again, user experience testing would be paramount in understanding the applicability of this augmentation.

Expanding the types of surfaces upon which the system works would also aid in the creation of more immersive interaction modalities with deep learning scratch input. Initial interest from Mann \textit{et al} \cite{mann_hyperacoustic} originated in part from work in water-based musical instrumentation. Perhaps deep learning scratch input could be applied similarly to other states of matter. Similarly, a deep learning approach to scratch input for musical expression may yield interesting results.

\section{Conclusion}

The extent to which deep learning can be used to enhance scratch input was investigated. A convolutional neural network-based system was proposed and tested, achieving a final testing accuracy of 95.8\% on a dataset collected from a variety of environments and surfaces using generic smartphones and tablets. These results reduced the gesture misclassification rate by 60\% as compared to the previous state-of-the-art \cite{harrison_scratch} despite the additional challenge associated with multiple environments and non-contact microphones. 

We concluded a high potential for deep learning in the creation of natural user interface (NUI) systems and other rapidly deployable user interfaces that make use of vibrational signals arising from user action (i.e., scratching, tapping, swiping). Home automation, accessibility, and hazardous environment use cases were highlighted, and challenges associated with the methodology in production were addressed. Next steps include further user testing, transfer learning for calibration, and experimentation with alternative deep learning models.

\section{Appendices}
\begin{table*}[h]
    \caption{Fixed and variable hyperparameters used in the CNN model}
    \label{tab:hyper-parameters}
    \centering
    \begin{tabular}{ |p{7em}|p{7em}|p{6.5em}|}
     \hline
     Fixed Hyperparameters & \multicolumn{2}{|c|}{Variable Hyperparameters (Name + Range)} \\
     \hline\hline
      2 Convolutional Layers &
      Input Image Transformation &
      Cropping/ padding/ squeezing/ stretching\\
     \hline
     2 Fully Connected Layers&
     Batch Size&
     [24, 48, 64, 126]\\
     \hline
     Batch Normalization&
     Number of Kernels in the Convolutional Layers&
     [4, 8, 12, 26]\\
     \hline
     Cross Entropy Loss function &
     Learning Rate&
     [0.3, 0.1, 0.01]\\
     \hline
     3x3 Kernel Size &
     Seed&
     [6, 42] \\
     \hline
     100 Neurons in the First Fully Connected Layer & & \\
     \hline
    \end{tabular}
\end{table*}

\begin{acks}
We acknowledge Prof. Jonathan Rose of the University of Toronto for his feedback and advice on this project. 
\end{acks}

\bibliographystyle{ACM-Reference-Format}
\bibliography{sample-base}


\begin{thebibliography}{20}


\ifx \showCODEN    \undefined \def \showCODEN     #1{\unskip}     \fi
\ifx \showDOI      \undefined \def \showDOI       #1{#1}\fi
\ifx \showISBNx    \undefined \def \showISBNx     #1{\unskip}     \fi
\ifx \showISBNxiii \undefined \def \showISBNxiii  #1{\unskip}     \fi
\ifx \showISSN     \undefined \def \showISSN      #1{\unskip}     \fi
\ifx \showLCCN     \undefined \def \showLCCN      #1{\unskip}     \fi
\ifx \shownote     \undefined \def \shownote      #1{#1}          \fi
\ifx \showarticletitle \undefined \def \showarticletitle #1{#1}   \fi
\ifx \showURL      \undefined \def \showURL       {\relax}        \fi
\providecommand\bibfield[2]{#2}
\providecommand\bibinfo[2]{#2}
\providecommand\natexlab[1]{#1}
\providecommand\showeprint[2][]{arXiv:#2}

\bibitem[\protect\citeauthoryear{Christine~Bizier}{Christine~Bizier}{2012}]%
        {disability_survey}
\bibfield{author}{\bibinfo{person}{Sabrina~Gilbert Christine~Bizier,
  Gail~Fawcett}.} \bibinfo{year}{2012}\natexlab{}.
\newblock \showarticletitle{Canadian Survey on Disability, 2012}.
\newblock \bibinfo{journal}{\emph{Statistics Canada}} (\bibinfo{year}{2012}).
\newblock


\bibitem[\protect\citeauthoryear{Fawcett}{Fawcett}{2006}]%
        {fawcett2006introduction}
\bibfield{author}{\bibinfo{person}{Tom Fawcett}.}
  \bibinfo{year}{2006}\natexlab{}.
\newblock \showarticletitle{An introduction to ROC analysis}.
\newblock \bibinfo{journal}{\emph{Pattern recognition letters}}
  \bibinfo{volume}{27}, \bibinfo{number}{8} (\bibinfo{year}{2006}),
  \bibinfo{pages}{861--874}.
\newblock


\bibitem[\protect\citeauthoryear{Goodfellow, Bengio, Courville, and
  Bengio}{Goodfellow et~al\mbox{.}}{2016}]%
        {goodfellow2016deep}
\bibfield{author}{\bibinfo{person}{Ian Goodfellow}, \bibinfo{person}{Yoshua
  Bengio}, \bibinfo{person}{Aaron Courville}, {and} \bibinfo{person}{Yoshua
  Bengio}.} \bibinfo{year}{2016}\natexlab{}.
\newblock \bibinfo{booktitle}{\emph{Deep learning}}. Vol.~\bibinfo{volume}{1}.
\newblock \bibinfo{publisher}{MIT press Cambridge}.
\newblock


\bibitem[\protect\citeauthoryear{Harrison and Hudson}{Harrison and
  Hudson}{2008}]%
        {harrison_scratch}
\bibfield{author}{\bibinfo{person}{Chris Harrison} {and}
  \bibinfo{person}{Scott~E. Hudson}.} \bibinfo{year}{2008}\natexlab{}.
\newblock \showarticletitle{Scratch Input: Creating Large, Inexpensive,
  Unpowered and Mobile Finger Input Surfaces}. In
  \bibinfo{booktitle}{\emph{Proceedings of the 21st Annual ACM Symposium on
  User Interface Software and Technology}} (Monterey, CA, USA)
  \emph{(\bibinfo{series}{UIST '08})}. \bibinfo{publisher}{Association for
  Computing Machinery}, \bibinfo{address}{New York, NY, USA},
  \bibinfo{pages}{205–208}.
\newblock
\showISBNx{9781595939753}
\urldef\tempurl%
\url{https://doi.org/10.1145/1449715.1449747}
\showDOI{\tempurl}


\bibitem[\protect\citeauthoryear{Harrison, Schwarz, and Hudson}{Harrison
  et~al\mbox{.}}{2011}]%
        {tapsense}
\bibfield{author}{\bibinfo{person}{Chris Harrison}, \bibinfo{person}{Julia
  Schwarz}, {and} \bibinfo{person}{Scott~E Hudson}.}
  \bibinfo{year}{2011}\natexlab{}.
\newblock \showarticletitle{TapSense: enhancing finger interaction on touch
  surfaces}. In \bibinfo{booktitle}{\emph{Proceedings of the 24th annual ACM
  symposium on User interface software and technology}}.
  \bibinfo{pages}{627--636}.
\newblock


\bibitem[\protect\citeauthoryear{Harrison, Tan, and Morris}{Harrison
  et~al\mbox{.}}{2010}]%
        {skinput}
\bibfield{author}{\bibinfo{person}{Chris Harrison}, \bibinfo{person}{Desney
  Tan}, {and} \bibinfo{person}{Dan Morris}.} \bibinfo{year}{2010}\natexlab{}.
\newblock \showarticletitle{Skinput: appropriating the body as an input
  surface}. In \bibinfo{booktitle}{\emph{Proceedings of the SIGCHI conference
  on human factors in computing systems}}. \bibinfo{pages}{453--462}.
\newblock


\bibitem[\protect\citeauthoryear{Hershey, Chaudhuri, Ellis, Gemmeke, Jansen,
  Moore, Plakal, Platt, Saurous, Seybold, Slaney, Weiss, and Wilson}{Hershey
  et~al\mbox{.}}{2017}]%
        {google_audio}
\bibfield{author}{\bibinfo{person}{Shawn Hershey}, \bibinfo{person}{Sourish
  Chaudhuri}, \bibinfo{person}{Daniel P.~W. Ellis}, \bibinfo{person}{Jort~F.
  Gemmeke}, \bibinfo{person}{Aren Jansen}, \bibinfo{person}{R.~Channing Moore},
  \bibinfo{person}{Manoj Plakal}, \bibinfo{person}{Devin Platt},
  \bibinfo{person}{Rif~A. Saurous}, \bibinfo{person}{Bryan Seybold},
  \bibinfo{person}{Malcolm Slaney}, \bibinfo{person}{Ron~J. Weiss}, {and}
  \bibinfo{person}{Kevin Wilson}.} \bibinfo{year}{2017}\natexlab{}.
\newblock \bibinfo{title}{CNN Architectures for Large-Scale Audio
  Classification}.
\newblock
\newblock
\showeprint[arxiv]{1609.09430}~[cs.SD]


\bibitem[\protect\citeauthoryear{Ioffe and Szegedy}{Ioffe and Szegedy}{2015}]%
        {ioffe2015batch}
\bibfield{author}{\bibinfo{person}{Sergey Ioffe} {and}
  \bibinfo{person}{Christian Szegedy}.} \bibinfo{year}{2015}\natexlab{}.
\newblock \showarticletitle{Batch normalization: Accelerating deep network
  training by reducing internal covariate shift}. In
  \bibinfo{booktitle}{\emph{International conference on machine learning}}.
  PMLR, \bibinfo{pages}{448--456}.
\newblock


\bibitem[\protect\citeauthoryear{Lei, Tu, Liu, Li, and Xie}{Lei
  et~al\mbox{.}}{2017}]%
        {voice_assistant_security}
\bibfield{author}{\bibinfo{person}{Xinyu Lei}, \bibinfo{person}{Guan-Hua Tu},
  \bibinfo{person}{Alex Liu}, \bibinfo{person}{Chiyu Li}, {and}
  \bibinfo{person}{Tian Xie}.} \bibinfo{year}{2017}\natexlab{}.
\newblock \showarticletitle{The Insecurity of Home Digital Voice Assistants -
  Amazon Alexa as a Case Study}.
\newblock  (\bibinfo{date}{12} \bibinfo{year}{2017}).
\newblock


\bibitem[\protect\citeauthoryear{Mann}{Mann}{2007}]%
        {mann_physiphone}
\bibfield{author}{\bibinfo{person}{Steve Mann}.}
  \bibinfo{year}{2007}\natexlab{}.
\newblock \showarticletitle{Natural Interfaces for Musical Expression:
  Physiphones and a Physics-Based Organology}. In
  \bibinfo{booktitle}{\emph{Proceedings of the 7th International Conference on
  New Interfaces for Musical Expression}} (New York, New York)
  \emph{(\bibinfo{series}{NIME '07})}. \bibinfo{publisher}{Association for
  Computing Machinery}, \bibinfo{address}{New York, NY, USA},
  \bibinfo{pages}{118–123}.
\newblock
\showISBNx{9781450378376}
\urldef\tempurl%
\url{https://doi.org/10.1145/1279740.1279761}
\showDOI{\tempurl}


\bibitem[\protect\citeauthoryear{Mann, Janzen, and Lo}{Mann
  et~al\mbox{.}}{2008}]%
        {mann_hyperacoustic}
\bibfield{author}{\bibinfo{person}{Steve Mann}, \bibinfo{person}{Ryan Janzen},
  {and} \bibinfo{person}{Raymond Chun~Hing Lo}.}
  \bibinfo{year}{2008}\natexlab{}.
\newblock \showarticletitle{Hyperacoustic instruments: Computer-controlled
  instruments that are not electrophones}. \bibinfo{pages}{89 -- 92}.
\newblock
\showISBNx{978-1-4244-2570-9}
\urldef\tempurl%
\url{https://doi.org/10.1109/ICME.2008.4607378}
\showDOI{\tempurl}


\bibitem[\protect\citeauthoryear{Ono, Shizuki, and Tanaka}{Ono
  et~al\mbox{.}}{2013}]%
        {touch_and_activate}
\bibfield{author}{\bibinfo{person}{Makoto Ono}, \bibinfo{person}{Buntarou
  Shizuki}, {and} \bibinfo{person}{Jiro Tanaka}.}
  \bibinfo{year}{2013}\natexlab{}.
\newblock \showarticletitle{Touch \& activate: adding interactivity to existing
  objects using active acoustic sensing}. In
  \bibinfo{booktitle}{\emph{Proceedings of the 26th annual ACM symposium on
  User interface software and technology}}. \bibinfo{pages}{31--40}.
\newblock


\bibitem[\protect\citeauthoryear{Palanisamy, Singhania, and Yao}{Palanisamy
  et~al\mbox{.}}{2020}]%
        {palanisamy2020rethinking}
\bibfield{author}{\bibinfo{person}{Kamalesh Palanisamy},
  \bibinfo{person}{Dipika Singhania}, {and} \bibinfo{person}{Angela Yao}.}
  \bibinfo{year}{2020}\natexlab{}.
\newblock \showarticletitle{Rethinking cnn models for audio classification}.
\newblock \bibinfo{journal}{\emph{arXiv preprint arXiv:2007.11154}}
  (\bibinfo{year}{2020}).
\newblock


\bibitem[\protect\citeauthoryear{Siegeltuch}{Siegeltuch}{2017}]%
        {water_icon}
\bibfield{author}{\bibinfo{person}{Mark Siegeltuch}.}
  \bibinfo{year}{2017}\natexlab{}.
\newblock \showarticletitle{The Water Symbol Its Origin and Transformation}.
\newblock  (\bibinfo{year}{2017}).
\newblock


\bibitem[\protect\citeauthoryear{{Tanveer}, {Zhu}, {Ahmed}, {Thomas}, {Imran},
  and {Salman}}{{Tanveer} et~al\mbox{.}}{2021}]%
        {9349416}
\bibfield{author}{\bibinfo{person}{M.~H. {Tanveer}}, \bibinfo{person}{H.
  {Zhu}}, \bibinfo{person}{W. {Ahmed}}, \bibinfo{person}{A. {Thomas}},
  \bibinfo{person}{B.~M. {Imran}}, {and} \bibinfo{person}{M. {Salman}}.}
  \bibinfo{year}{2021}\natexlab{}.
\newblock \showarticletitle{Mel-spectrogram and Deep CNN Based Representation
  Learning from Bio-Sonar Implementation on UAVs}. In
  \bibinfo{booktitle}{\emph{2021 International Conference on Computer, Control
  and Robotics (ICCCR)}}. \bibinfo{pages}{220--224}.
\newblock
\urldef\tempurl%
\url{https://doi.org/10.1109/ICCCR49711.2021.9349416}
\showDOI{\tempurl}


\bibitem[\protect\citeauthoryear{Thornton}{Thornton}{[n.d.]}]%
        {thorntonaudio}
\bibfield{author}{\bibinfo{person}{Boyang Zhang Jared Leitner~Sam Thornton}.}
  \bibinfo{year}{[n.d.]}\natexlab{}.
\newblock \showarticletitle{Audio Recognition using Mel Spectrograms and
  Convolution Neural Networks}.
\newblock  (\bibinfo{year}{[n.\,d.]}).
\newblock


\bibitem[\protect\citeauthoryear{Villanueva and Dr{\"o}gehorn}{Villanueva and
  Dr{\"o}gehorn}{2018}]%
        {home_auto_gestures}
\bibfield{author}{\bibinfo{person}{Marcel Villanueva} {and}
  \bibinfo{person}{Olaf Dr{\"o}gehorn}.} \bibinfo{year}{2018}\natexlab{}.
\newblock \showarticletitle{{Using Gestures to Interact with Home Automation
  Systems: A Socio-Technical Study on Motion Capture Technologies for Smart
  Homes}}. In \bibinfo{booktitle}{\emph{{SEEDS International Conference -
  Sustainable, Ecological, Engineering Design for Society}}}.
  \bibinfo{address}{Dublin, Ireland}.
\newblock
\urldef\tempurl%
\url{https://hal.archives-ouvertes.fr/hal-02179898}
\showURL{%
\tempurl}


\bibitem[\protect\citeauthoryear{Wyse}{Wyse}{2017}]%
        {audio_overfit}
\bibfield{author}{\bibinfo{person}{L. Wyse}.} \bibinfo{year}{2017}\natexlab{}.
\newblock \bibinfo{title}{Audio Spectrogram Representations for Processing with
  Convolutional Neural Networks}.
\newblock
\newblock
\showeprint[arxiv]{1706.09559}~[cs.SD]


\bibitem[\protect\citeauthoryear{Yaser, Malik, and Hsuan-Tien}{Yaser
  et~al\mbox{.}}{2012}]%
        {yaser2012learning}
\bibfield{author}{\bibinfo{person}{S~Abu-Mostafa Yaser},
  \bibinfo{person}{Magdon-Ismail Malik}, {and} \bibinfo{person}{Lin
  Hsuan-Tien}.} \bibinfo{year}{2012}\natexlab{}.
\newblock \showarticletitle{Learning from data}.
\newblock \bibinfo{journal}{\emph{Renesselaer Polytechnic Institute}}
  (\bibinfo{year}{2012}).
\newblock


\bibitem[\protect\citeauthoryear{Zwicker}{Zwicker}{1961}]%
        {zwicker1961subdivision}
\bibfield{author}{\bibinfo{person}{Eberhard Zwicker}.}
  \bibinfo{year}{1961}\natexlab{}.
\newblock \showarticletitle{Subdivision of the audible frequency range into
  critical bands (Frequenzgruppen)}.
\newblock \bibinfo{journal}{\emph{The Journal of the Acoustical Society of
  America}} \bibinfo{volume}{33}, \bibinfo{number}{2} (\bibinfo{year}{1961}),
  \bibinfo{pages}{248--248}.
\newblock


\end{thebibliography}

\end{document}